\documentclass[a4paper,12pt]{article}

\begin{document}

\author{G. F. Bressange\thanks{E-mail : georges.bressange@ucd.ie} \, and 
P.A. Hogan\thanks{E-mail : phogan@ollamh.ucd.ie}\\
\small Mathematical Physics Department\\
\small  University College Dublin, Belfield, Dublin 4, Ireland}

\title{Peeling Properties of Light--Like Signals in 
General Relativity}
\date{}
\maketitle

\begin{abstract}
The peeling properties of a light--like signal propagating 
through a general Bondi--Sachs vacuum space--time and 
leaving behind another Bondi--Sachs vacuum space--time are 
studied. We demonstrate that in general the peeling behavior is 
the conventional one which is associated with a radiating 
isolated system and that it becomes unconventional if the 
asymptotically flat space--times on either side of the 
history of the light--like signal tend to flatness at 
future null infinity faster than the general Bondi--Sachs 
space--time. This latter situation occurs if, for example, 
the space--times in question are static Bondi--Sachs space--
times.

\end{abstract}
\thispagestyle{empty}
\newpage

\section{Introduction}\indent
The history of a light--like signal in General Relativity is 
a singular null hypersurface. The null hypersurface is 
called singular because in general the Ricci tensor and 
the Weyl tensor of the space--time contain Dirac $\delta -$
function terms with the $\delta -$function singular on the 
null hypersurface. Such a singular null hypersurface 
can be used as a simplified model of a supernova \cite{BBH} 
if the space--time before and after the emission of the 
light--like signal is a model of the vacuum field due to 
an isolated gravitating system and if the singular null 
hypersurface is asymptotically (as future null infinity is 
approached) a future--directed null--cone. For the model 
described in \cite{BBH} the space--times 
before and after the emission of the light--like signal 
are two copies of the Weyl asymptotically flat, static 
space--times \cite{KSHM} but with different 
multipole moments. Thus 
the explosion is modelled by a sudden change in the 
multipole moments of the source. The coefficients of the $\delta -$
function in the Weyl tensor display unconventional 
peeling properties. By peeling properties here we mean 
the dependence of the coefficients of the $\delta -$function 
in the Weyl tensor on an affine parameter $r$ along the 
generators of the singular null hypersurface, with $r\longrightarrow 
+\infty$ as future null infinity is approached. In general the $\delta -$
function in the Weyl tensor can be unambiguously split into a 
matter part (if it exists) of Petrov type II and a wave 
part (if it exists) of Petrov type N. This splitting, 
which was originally announced in \cite{BI}, is fully developed 
in \cite{BBH}. In the simplified supernova model described 
in \cite{BBH} the matter part of the $\delta -$function 
in the Weyl tensor depends on $r$ in the form of $O(r^{-3})$--
terms and smaller terms, while the wave part of the $\delta -$
function in the Weyl tensor has a coefficient which is 
$O(r^{-4})$. We denote the components of the $\delta$--function in the Weyl tensor, in Newman--Penrose notation, 
by $\hat\Psi _A\ (A=0, 1, 2, 3, 4)$ chosen in such a way  
that the first four of these describe the matter part of 
the Weyl tensor and $\hat\Psi _4$ describes the 
wave part. Defining $\hat\Psi _A$ in this way the conventional 
peeling behavior would be to have $\hat\Psi _A=O(
r^{-5+A})\ (A=0, 1, 2, 3, 4)$. This is the conventional peeling behavior 
when compared with models of vacuum gravitational fields 
due to isolated gravitating systems in general (cf. 
\cite{RT1}--\cite{NP} and (2.20)--(2.24) below). The purpose of the present paper is 
to explain the unconventional peeling behavior described 
above by putting 
it in the context of light--like signals emitted by 
a general class of isolated gravitating systems. 

We take the vacuum gravitational field outside an 
isolated system, before and after the emission of a 
light--like signal, to be modelled by a Bondi--Sachs \cite{BVBM} 
\cite{S} asymptotically flat space--time. We also consider 
the most general matching of the two space--times on the 
singular null hypersurface separating them. How the two 
space--times are glued together influences the type of 
signal whose history is the singular null hypersurface. 
We demonstrate that in general the peeling behavior is 
the conventional one which is associated with a radiating 
isolated system and that it becomes unconventional if the 
asymptotically flat space--times on either side of the 
history of the light--like signal tend to flatness at 
future null infinity faster than the general Bondi--Sachs 
space--time.

\setcounter{equation}{0}
\section{Light--Like Signal From Isolated Source}\indent
Throughout this paper the space--time model of the vacuum 
gravitational field outside an isolated source will 
be a Bondi--Sachs \cite{BVBM}, \cite{S} space--time. A 
convenient form of this space--time line--element is 
given by \cite{HT}
\begin{equation}\label{2.1}
ds^2=(\theta ^1)^2+(\theta ^2)^2-2\,\theta ^3\theta ^4\ ,
\end{equation}
with 
\begin{eqnarray}
\theta ^1&=&rp^{-1}({\rm e}^{\alpha}\cosh\beta\,dx+{
\rm e}^{-\alpha}\sinh\beta\,dy+a\,du)\ ,\\\label{2.2}
\theta ^2&=&rp^{-1}({\rm e}^{\alpha}\sinh
\beta\,dx+{
\rm e}^{-\alpha}\cosh\beta\,dy+b\,du)
\ ,\\\label{2.3}
\theta ^3&=&-dr-\frac{1}{2}c\,du\ ,\\\label{2.4}
\theta ^4&=&-du\ .
\end{eqnarray}
The six function $\alpha , \beta , a, b, p, 
c$ depend on all coordinates $x, y, r, u$ and 
$u={\rm const.}$ are null hypersurfaces 
generated by the geodesic integral curves 
of the vector field $\frac{\partial}{\partial 
r}$ with $r$ an affine parameter along these 
geodesics. The following assumptions are made regarding 
the $r$--dependence of these functions:
\begin{eqnarray}
\alpha &=&\frac{\alpha _1}{r}+\frac{\alpha _2}
{r^2}+\frac{\alpha _3}{r^3}+\dots\ ,\\\label{2.6}
\beta &=&\frac{\beta _1}{r}+\frac{\beta _2}
{r^2}+\frac{\beta _3}{r^3}+\dots\ ,
\\\label{2.7}
a&=&a_0+\frac{a_1}{r}+\frac{a_2}{r^2}+\frac
{a_3}{r^3}+\dots\ ,\\\label{2.8}
b&=&b_0+\frac{b_1}{r}+\frac{b_2}{r^2}+\frac
{b_3}{r^3}+\dots\ ,\\\label{2.9}
p&=&p_0\left (1+\frac{q_1}{r}+\frac{q_2}{r^2}+
\frac{q_3}{r^3}+\dots\right )\ ,\\\label{2.10}
c&=&1-\frac{2\,m}{r}+\dots\ .
\end{eqnarray}
Here $p_0=1+\frac{1}{4}(x^2+y^2)$ and 
all the other coefficients of the inverse 
powers of $r$ displayed above are functions 
of $(x, y, u)$. All eighteen functions of 
$(x, y, u)$ appearing in (2.6)--(2.11) 
are required in order to have a knowledge 
of the metric tensor components up to and 
including $\frac{1}{r}$--terms. The vacuum 
field equations and an outgoing radiation 
condition \cite{HT} allow us to specialise 
the sixteen functions as follows:
\begin{equation}\label{2.12}
\alpha _2=\beta _2=0\ ,
\end{equation}
\begin{equation}\label{2.13}
a_0=a_1=0\ ,\qquad {\rm and}\qquad b_0=b_1=0\ ,
\end{equation}
\begin{eqnarray}
a_2&=&p_0^4\left\{\frac{\partial}{\partial x}(p_0^{-2}\alpha _1)
+\frac{\partial}{\partial y}(p_0^{-2}\beta _1)\right\}\ ,\\\label{2.14}
b_2&=&p_0^4\left\{\frac{\partial}{\partial x}(p_0^{-2}\beta _1)
-\frac{\partial}{\partial y}(p_0^{-2}\alpha _1)\right\}\ ,
\end{eqnarray}
\begin{equation}\label{2.16}
q_1=0\ ,\qquad q_2=\frac{1}{2}(\alpha _1^2+\beta _1^2)\ ,\qquad 
q_3=0\ .
\end{equation} 
When these equations are satisfied there remain five further field 
equations. They are propagation equations for $m(u, x, y), \alpha _3
(u, x, y), \beta _3 (u, x, y), a_3(u, x, y)$ and $b_3(u, x, y)$ off 
$u={\rm const}$. The simplest reads
\begin{equation}\label{2.17}
\dot M+\left |\dot\gamma\right |^2=0\ ,
\end{equation}
where the dot denotes partial differentiation with respect to $u$, and 
\begin{equation}\label{2.18}
M=m-\dot q_2-\frac{1}{2}\left\{\frac{\partial}{\partial x}(p_0^{-2}
a_2)+\frac{\partial}{\partial y}(p_0^{-2}b_2)\right \}\ ,
\end{equation}
with
\begin{equation}\label{2.19}
\gamma =\alpha _1+i\beta _1\ .
\end{equation}
When the equation (\ref{2.17}) is averaged over the 2--sphere with 
line--element $dl^2=p_0^{-2}(dx^2+dy^2)$ the well--known Bondi--Sachs 
mass--loss formula results. The remaining equations involving $\dot \alpha _3, 
\dot\beta _3, \dot a_3$ and $\dot b_3$ are given in \cite{HT} and 
will not be used here. The curvature tensor components, in Newman--
Penrose notation, for the space--time described above display the 
conventional peeling behavior:
\begin{eqnarray}
\Psi_0&=&-\frac{1}{r^5}\left\{6(\alpha _3+i\beta _3)-\frac{3}{2}
(\gamma +\bar\gamma )^2(\gamma -\bar\gamma )-2\bar\gamma ^3\right\}+\dots\ 
,\\\label{2.20}
\Psi_1&=&-\frac{1}{r^4\sqrt{2}}\left\{\frac{3}{2}p_0^{-1}(a_3+ib_3)+3p_0^3\gamma\,
\frac{\partial}{\partial\bar z}(p_0^{-2}\bar\gamma)\right\}+\dots\ ,
\\\label{2.21}
\Psi_2&=&-\frac{1}{r^3}\left\{M+\gamma\,\frac{\partial\bar\gamma}
{\partial u}+2p_0^2\frac{\partial}{\partial\bar z}\left (\frac{\partial}
{\partial\bar z}(p_0^{-2}\bar\gamma )\right )\right\}+\dots\ ,\\\label{2.22}
\Psi _3&=&-\frac{2}{r^2\sqrt{2}}p_0^2\frac{\partial}{\partial u}\left (
p_0\frac{\partial}{\partial\bar z}(p_0^{-2}\bar\gamma )\right )+\dots\ ,\\\label{2.23}
\Psi_4&=&-\frac{1}{r}\,\frac{\partial ^2\bar\gamma}{\partial u^2}+\dots\ ,
\end{eqnarray}
where we have put $z=x+iy$ and a bar denotes complex conjugation. Finally 
the complex shear $\sigma$ and the real expansion $\vartheta$ of the 
null geodesic integral curves of the vector field $\frac{\partial}{\partial 
r}$ are given by
\begin{eqnarray}
\sigma &=&-\frac{(\alpha _1+i\beta _1)}{r^2}-\frac{3(\alpha _3+i\beta _3)+
2\alpha _1\beta _1^2}{r^4}+O\left (r^{-5}\right )\ ,\\\label{2.26}
\vartheta &=&\frac{1}{r}+\frac{2q_2}{r^3}+O\left (r^{-5}\right )\ ,
\end{eqnarray}
respectively, demonstrating that asymptotically (as $r\longrightarrow +\infty$) 
the null hypersurfaces $u={\rm const.}$ are future--directed null--cones.

We now consider the space--time above to be subdivided into two halves $M^-(u\leq 0)$ 
and $M^+(u\geq 0)$ each with boundary the null hypersurface $u=0$. Let 
$x^\mu _+=(x_+, y_+, r_+, u)$ (with $\mu =1, 2, 3, 4$) be the local coordinate system in $M^+$ 
in which the line--element takes the form given by (\ref{2.1})--(\ref{2.4}) 
with the coefficients of the powers of $r_+^{-1}$ in (2.6)--(2.11) 
denoted with a superscript plus as $\alpha _1^+, \alpha _2^+$, etc.. Let 
$x^\mu =(x, y, r, u)$ be the local coordinate system in $M^-$ in terms of 
which the line--element of the space--time has the form given by (2.1)
--(2.5). We take $\xi ^a=(x, y, r)$, with $a=1, 2, 3$, as local intrinsic coordinates 
on $u=0$. Now $u=0$ is the history of a light--like signal emitted by 
the isolated source and propagating into the space--time $M^-$ leaving the 
space--time $M^+$ behind. We apply the Barrab\`es--Israel \cite{BI} 
technique  (see also \cite{BH} for some recent developments) to analyse 
the physical properties of the signal with history $u=0$. We shall assume 
that the reader is familiar with \cite{BI}. While we are applying 
the technique to the current situation we will comment on what we are 
doing so as to guide the reader by example through \cite{BI}. The 
first requirement of \cite{BI} is that the metric tensors induced on the 
null hypersurface $u=0$ by its embedding in $M^+$ and in $M^-$ agree on 
$u=0$. This is achieved if the space--times $M^+$ and $M^-$ described 
above are attached on $u=0$ with the following matching conditions:
\begin{eqnarray}\label{2.27}
x_+&=&f(x, y)+\frac{f_1(x, y)}{r}+O\left (r^{-2}\right )\ ,\\\label{2.28}
y_+&=&g(x, y)+\frac{g_1(x, y)}{r}+O\left (r^{-2}\right )\ ,\\\label{2.29}
r_+&=&r\,h(x, y)+h_0(x, y)+O\left (r^{-1}\right )\ ,
\end{eqnarray}
with 
\begin{equation}\label{2.31}
f_x=g_y\ ,\qquad f_y=-g_x\ ,\qquad h=\frac{P}{p_0}\left\{f_x^2+g_x^2\right\}
^{-\frac{1}{2}}\ ,
\end{equation}
where $P=1+\frac{1}{4}(f^2+g^2)$ and the subscripts on $f, g$ denote 
partial differentiation. In addition we must have 
\begin{eqnarray}\lefteqn{
hP^{-2}(f_x^2-f_y^2)\,\beta _1^++2\,hP^{-2}f_x\,f_y\,\alpha _1
^+-p_0^{-2}\beta _1=}\nonumber\\
& & -\frac{1}{2}h^2P^{-2}\left\{f_x\,
\left (\frac{\partial g_1}{\partial x}+\frac{\partial f_1}{\partial 
y}\right )+f_y\,\left (\frac{\partial f_1}{\partial x}-\frac{
\partial g_1}{\partial y}\right )\right\}\ ,
\end{eqnarray}
\begin{eqnarray}\lefteqn{
hP^{-2}(f_x^2-f_y^2)\,\alpha _1^+-2\,hP^{-2}f_x\,f_y\,\beta _1
^+-p_0^{-2}\alpha _1=}\nonumber\\
& & -\frac{1}{2}h^2P^{-2}\left\{f_x\,
\left (\frac{\partial f_1}{\partial x}-\frac{\partial g_1}
{\partial 
y}\right )-f_y\,\left (\frac{\partial g_1}{\partial x}+\frac{
\partial f_1}{\partial y}\right )\right\}\ ,
\end{eqnarray}
and
\begin{eqnarray}\lefteqn{
-h_0+\frac{1}{2}p_0^{-1}(f_x^2+f_y^2)^{-\frac{1}{2}}(f\,f_1+
g\,g_1)=}\nonumber\\
& & \frac{1}{2}P\,p_0^{-1}(f_x^2+f_y^2)^{-\frac{3}{2}}
\left\{f_x\,
\left (\frac{\partial f_1}{\partial x}+\frac{\partial g_1}
{\partial 
y}\right )+f_y\,\left (\frac{\partial f_1}{\partial y}-\frac{
\partial g_1}{\partial x}\right )\right\}\ .
\end{eqnarray}
Here $\beta _1^+=\beta _1^+(f, g, 0), \beta _1=
\beta _1(x, y, 0), \alpha _1^+=\alpha _1^+(f, g, 0)$ 
and $\alpha _1=\alpha _1(x, y, 0)$. 
The complexity of these matching conditions suggests that 
we examine the light--like signal in two stages. The 
leading terms in (2.27)--(2.28) are constructed from the 
analytic function $F(z)=f(x, y)+ig(x, y)$ with $z=x+iy$. 
They describe a part of the gluing of $M^+$ to $M^-$ on 
$u=0$ which is a Penrose warp \cite{P}. This particular gluing 
leads to an impulsive gravitational wave with history $u=0$ having 
a line or directional singularity and we will consider it in 
section 3. Thus for the remainder of this section we shall 
take $f(x, y)=x$ and $g(x, y)=y$ in (2.27) and (2.28) and 
thus in (2.29) $h=1$. Now (2.31)--(2.33) simplify to read 
\begin{equation}\label{2.34}
\left [\beta _1\right ]=-\frac{1}{2}\left (\frac{\partial g_1}{\partial 
x}+\frac{\partial f_1}{\partial y}\right )\ ,
\qquad \left [\alpha _1\right ]=-\frac{1}{2}\left 
(\frac{\partial f_1}{\partial 
x}-\frac{\partial g_1}{\partial y}\right )\ ,
\end{equation}
and
\begin{equation}\label{2.35}
h_0=-\frac{1}{2}p_0^2\frac{\partial}{\partial x}(p_0
^{-2}f_1)-\frac{1}{2}p_0^2\frac{\partial}{\partial y}(p_0
^{-2}g_1)\ ,
\end{equation}
where the square brackets will henceforth denote the jump 
across $u=0$ of the quantity within them, calculated in the 
coordinates $x^\mu _-$. Thus, for example, $\left [\beta _1\right ]=
\beta _1^+(x, y, u=0)-\beta _1(x, y, u=0)$.

The Barrab\`es--Israel technique is an extension to hypersurfaces 
of all types of the extrinsic curvature technique for non--null 
hypersurfaces (see, for example \cite{I}). In the case of a null 
hypersurface the normal is tangent to the hypersurface so in 
order to obtain an analogous quantity to extrinsic curvature one 
first constructs a `transverse vector field' which is any vector 
field defined on the hypersurface which is not tangent to the 
hypersurface and which is the same vector field when viewed in 
the coordinates $(x^\mu _+)$ and in the coordinates $(x^\mu _-)$. 
In our case the normal to $u=0$ is given via the 1--form
\begin{equation}\label{2.36}
n_\mu\,dx^\mu |_{\pm}=-du\ ,
\end{equation}
where $|_{\pm}$ means the quantity is calculated in the 
plus or minus coordinates. A natural choice of transversal on 
the minus side, in view of the form of the line--element given 
by (2.1)--(2.5), is
\begin{equation}\label{2.37}
{}^-N_\mu\,dx^\mu_-=-dr-\frac{1}{2}\,c\,du\ ,
\end{equation}
with $c$ given by (2.11). To ensure that the transversal 
when viewed on the plus side, ${}^+N_\mu$, is the same covariant 
vector field as ${}^-N_\mu$ we proceed as follows: We pointed 
out above that we may use $\xi ^a=(x, y, r)$ as intrinsic 
coordinates on $u=0$. We then have three linearly independent 
tangent vectors to $u=0$ given by $\frac{\partial}{\partial\xi 
^a}$. On the minus side of $u=0$ these have components 
\begin{equation}\label{2.38}
e^\mu _{(a)}|_-=\frac{\partial x^\mu _-}{\partial\xi ^a}=
\delta ^\mu _a\ ,
\end{equation}
and on the plus side their components are
\begin{equation}\label{2.39}
e^\mu _{(a)}|_+=\frac{\partial x^\mu _+}{\partial\xi ^a}\ ,
\end{equation}
with $x^\mu _+$ given in terms of $\xi ^a$ by (2.27)--(2.29) 
[now with $f=x, g=y, h=1$]. Now ${}^+N_\mu$ is chosen so 
that 
\begin{equation}\label{2.40}
\left [N_\mu\,e^\mu _{(a)}\right ]=0=\left [N_\mu\,N^\mu\right ]\ .
\end{equation}
With ${}^+N_\mu$ thus calculated the `transverse extrinsic 
curvature' on the plus or minus sides of $u=0$ is defined by
\begin{equation}\label{2.41}
{\cal K}^{\pm}_{ab}=-{}^{\pm}N_\mu\,\left (\frac{\partial 
e^\mu _{(a)}|_{\pm}}{\partial\xi ^b}+
{}^{\pm}\Gamma ^\mu _{\lambda\sigma}\,e^\lambda _{(a)}
|_{\pm}e^\sigma _{(b)}|_{\pm}\right )={\cal K}^{\pm}_{ba}\ ,
\end{equation}
where ${}^{\pm}\Gamma ^\mu _{\lambda\sigma}$ are the components 
of the Riemannian connection calculated on the plus or minus 
sides of $u=0$. We define 
\begin{equation}\label{2.42}
\gamma _{ab}=2\left [{\cal K}_{ab}\right ]\ ,
\end{equation}
and this is independent of the choice of transversal \cite{BI}. 
Now $\gamma _{ab}$ is extended to a 4--tensor $\gamma _{\mu\nu}$ 
on $u=0$ by padding out with zeros (the only requirement on 
the extension being that its projection tangential to $u=0$ be 
$\gamma _{ab}$). With 
\begin{equation}\label{2.43}
\gamma ^\mu=\gamma ^{\mu\nu}n_\nu\ ,\qquad \gamma ^{\dagger}
=\gamma ^\mu n_{\mu}\ ,\qquad \gamma =g^{\mu\nu}\gamma _{\mu\nu}\ 
,
\end{equation}
calculated in the plus or minus coordinates (we leave out the 
designation $|_{\pm}$ in such situations), the coefficient of 
the delta function $\delta (u)$ in the Einstein tensor of the 
space--time $M^+\cup M^-$ gives the surface stress--energy 
tensor \cite{BI}
\begin{equation}\label{2.44}
16\pi\,\eta ^{-1}S^{\mu\nu}=2\gamma ^{(\mu}n^{\nu)}-\gamma\,
n^\mu\,n^\nu-\gamma ^{\dagger}g^{\mu\nu}\ ,
\end{equation}
where $\eta ^{-1}=n^\mu\,N_\mu$. The coefficient of the delta 
function $\delta (u)$ in the Weyl tensor components is then 
\cite {BI}
\begin{equation}\label{2.45}
\hat C^{\kappa\lambda}{}_{\mu\nu}=2\eta\,n^{[\kappa}
\gamma ^{\lambda ]}{}_{[\mu}\,n_{\nu ]}-16\pi\,\delta ^{[\kappa}
_{[\mu}\,S^{\lambda ]}_{\nu ]}+\frac{8\pi}{3}\,
S^{\alpha}_{\alpha}\,\delta ^{\kappa\lambda}_{\mu\nu}\ .
\end{equation}
If $m^{\mu}$ is a unit complex vector field defined on $u=0$ 
which is tangential to $u=0$ and also orthogonal 
to the transversal then the Newman--Penrose components of 
$\hat C^{\kappa\lambda}{}_{\mu\nu}$ are given by \cite{BH}
\begin{eqnarray}\label{2.46}
\lefteqn{\hat\Psi _0=0\ ,\qquad \hat\Psi _1=0\ ,\qquad \hat\Psi _2=
-\frac{1}{6}\eta\,\gamma ^{\dagger}\ ,}\nonumber\\
& & \hat\Psi _3=-\frac{1}{2}\eta\,\gamma _\mu\,\bar m^\mu\ ,
\qquad \hat\Psi _4=-\frac{1}{2}\eta\,\gamma _{\mu\nu}\,\bar m
 ^\mu\,\bar m ^\nu\ .
\end{eqnarray}
In general the signal is Petrov type II and contains a gravitational 
wave if $\hat\Psi _4\neq 0$.

Carrying out this procedure (the calculations in this paper 
have been performed using GRTensorM version 1.2 for {\it MATHEMATICA} 3.x
\cite{Wolf} ) we find 
that $u=0$ has a non--vanishing surface stress--energy tensor 
with components
\begin{eqnarray}
S^{11}&=&O\left (\frac{1}{r^5}\right )\ ,\\
S^{22}&=&O\left (\frac{1}{r^5}\right )\ ,\\
S^{12}&=&O\left (\frac{1}{r^6}\right )\ ,\\
S^{13}&=&\frac{1}{16\pi\,r^3}\left\{2\,[a_2]+f_1-2p_0^2\frac{\partial h_0}{\partial x}
-2f_1\dot\alpha _1^+-2g_1\dot\beta _1^+\right\}+\dots\ ,\\
S^{23}&=&\frac{1}{16\pi\,r^3}\left\{2\,[b_2]+g_1-2p_0^2
\frac{\partial h_0}{\partial y}
-2f_1\dot\beta _1^++2g_1\dot\alpha _1^+\right\}+\dots\ ,
\end{eqnarray}
and
\begin{equation}\label{2.52}
S^{33}=-\frac{1}{4\pi\,r^2}\left\{[m-\dot q_2]-\frac{1}{2}p_0^2\frac{\partial
}{\partial y}(p_0^{-2}[b_2])-\frac{1}{2}p_0^2\frac{\partial
}{\partial x}(p_0^{-2}[a_2])+\frac{1}{2}(h_0+\Delta h_0)\right \}+\dots\ ,
\end{equation}
where $\Delta =p_0^2\left (\frac{\partial ^2}{\partial x^2}+
\frac{\partial ^2}{\partial y^2}\right )$ and $q_2$ is 
given by (2.16). The jumps $[a_2], [b_2]$ 
can be written in terms of the jumps $[\alpha _1], [\beta _1]$ using 
the field equations (2.14) and (2.15) and thence in terms of the 
functions $f_1, g_1$ via (2.34) to arrive at
\begin{eqnarray} 
\left[a_2\right]&=&-p_0\frac{\partial p_0}{\partial x}\left (\frac{\partial g_1}{\partial y}
-\frac{\partial f_1}{\partial x}\right )
+p_0\frac{\partial p_0}{\partial y}\left (\frac{\partial g_1}{\partial x}
+\frac{\partial f_1}{\partial y}\right ) -\frac{1}{2}\,\Delta f_1\ ,\\ 
\left[b_2\right]&=&p_0\frac{\partial p_0}{\partial x}\left (\frac{\partial g_1}{\partial x}
+\frac{\partial f_1}{\partial y}\right )
+p_0\frac{\partial p_0}{\partial y}\left (\frac{\partial g_1}{\partial y}
-\frac{\partial f_1}{\partial x}\right ) -\frac{1}{2}\,\Delta g_1\ .
\end{eqnarray}
and finally at
\begin{equation}
S^{33}=-\frac{1}{4\pi\,r^2}\left\{[m-\dot q_2]+\frac{1}{4}\,p_0^2\,
		\left ( \frac{\partial}{\partial x}(p_0^{-2}f_1)
			+ \frac{\partial}{\partial y}(p_0^{-2}g_1)
			\right )
		\right \}+\dots\ .
\end{equation}
The surface energy--density of 
the shell measured by a radially moving observer (see \cite{BI}) 
is a positive multiple of $S^{33}$. If we make the 
assumption that the functions $p_0^{-2}f_1, p_0^{-2}g_1$, defined on 
the 2--sphere with line--element $dl^2=p_0^{-2}(dx^2+dy^2)$, are 
bounded then it follows from (2.55) that {\it the leading term in 
the surface energy density averaged over the 
2--sphere is proportional to the jump in the Bondi--Sachs mass 
across $u=0$} (this latter, as mentioned following (2.19), is 
the average of $M$ in (2.18) over the 2--sphere). The average 
surface energy--density is positive if there is a {\it loss} 
of Bondi--Sachs mass.

Finally we calculate $\hat\Psi _A$ for $A=2, 3, 4$ in (2.46). The 
result can be put in the form
\begin{eqnarray}
\hat\Psi _2&=&O\left (\frac{1}{r^3}\right )\ ,\\
\hat\Psi _3&=&-4\pi\sqrt{2}\,r\,p_0^{-1}\left (S^{13}-iS^{23}\right )+
O\left (\frac{1}{r^3}\right )=O\left (\frac{1}{r^2}\right )\ ,
\end{eqnarray}
and
\begin{equation}\label{2.55}
\hat\Psi _4=-\frac{[\dot\alpha _1-i\dot\beta _1]}{r}+\frac{W}{r^2}+
O\left (\frac{1}{r^3}\right )\ ,
\end{equation}
with
\begin{eqnarray}\label{2.56}
\lefteqn{
W=2\frac{\partial}{\partial z}\left (p_0^2\frac{\partial h_0}{
\partial z}\right )-\frac{1}{2}[\alpha _1-i\beta _1]+\frac{\partial}{
\partial z}[a_2-i b_2]}\nonumber\\
& & \hspace{1cm}+(\dot\alpha _1^+
-i\dot\beta _1^+)\left \{h_0-i\left (\frac{\partial g_1}{\partial x}
-\frac{\partial f_1}{\partial y}\right )\right\}\ .
\end{eqnarray}
From this we arrive at the main result of this section, namely, {\it 
in general the coefficients of the delta function terms in the Weyl tensor display 
the conventional peeling behavior}. The signal contains an impulsive gravitational 
wave part $\hat\Psi _4$ with the expected $\frac{1}{r}$--behavior 
provided $[\dot\alpha _1-i\dot\beta _1]\neq 0$. This latter means that 
there is a jump in the Bondi `news' across $u=0$. This impulsive wave is 
accompanied by a light--like shell with surface stress--energy tensor 
given by (2.47)--(2.55). We see from (2.58) and (2.59) that if the matching 
(2.27)--(2.29) is the identity matching then the wave produced by the jump 
in the news across $u=0$ is free from line singularities.

\setcounter{equation}{0}
\section{A General and a Special Example}\indent
We return now to the general matching conditions (2.27)--(2.33). 
A considerable computational effort is needed to establish the following 
orders of magnitude of the components of the stress--energy tensor 
on $u=0$ in this case:
\begin{equation}\label{3.1}
S^{11}=O\left (\frac{1}{r^5}\right )\ ,\qquad S^{22}=
O\left (\frac{1}{r^5}\right )\ ,\qquad S^{12}=
O\left (\frac{1}{r^6}\right )\ ,
\end{equation}
\begin{equation}\label{3.2}
S^{13}=O\left (\frac{1}{r^3}\right )\ ,\qquad S^{23}=
O\left (\frac{1}{r^3}\right )\ ,\qquad S^{33}=
O\left (\frac{1}{r^2}\right )\ .
\end{equation}
In addition 
$\hat\Psi _A$ for $A=2, 3, 4$ in this case satisfy 
\begin{equation}\label{3.3}
\hat\Psi _2=O\left (\frac{1}{r^3}\right )\ ,\qquad 
\hat\Psi _3=O\left (\frac{1}{r^2}\right )\ ,
\end{equation}
and
\begin{equation}\label{3.4}
\hat\Psi _4=\frac{1}{r}\left\{p_0^2\left( H(z)-\frac{(F')^2}{1+{1\over 4}|F(z)|^2} (\dot\alpha _1^+
-i\dot\beta _1^+)\right)+\dot\alpha _1-i\dot\beta _1\right\}+
O\left (\frac{1}{r^2}\right )\ .
\end{equation}
Here, as in the paragraph following (2.33), $F(z)=f(x, y)+ig(x, y)$ 
with $z=x+iy$, $F'=dF/dz$ and
\begin{equation}\label{3.5}
H(z)=\frac{F'''}{F'}-\frac{3}{2}\left (\frac{F''}{F}\right )^2\ .
\end{equation}
Clearly when $F(z)=z$ this reduces to (2.58). We see from (3.3) 
and (3.4) that {\it in this general case the conventional peeling 
behavior is exhibited}. However we also see from the first $\frac{1}{r}$--
term in (3.4) that the Penrose 
wave has a directional or line singularity (as $z\bar z\longrightarrow 
+\infty$) \cite{P}.

We mentioned in the introduction that the principal motivation 
for the present study is to put into perspective the unconventional peeling 
behavior of $\hat\Psi _A$ for $A=2, 3, 4$ encountered in a simple 
example of a supernova \cite{BBH} in which the space--times $M^+$ and 
$M^-$ are two copies of the Weyl asymptotically flat, static space--
times having different multipole moments. We shall now demonstrate 
how this example emerges as a special case of the general situation 
described in section 2.

An asymptotically flat Weyl static space--time has a line--element
which can be put in the form (2.1)--(2.11) with 
\begin{equation}\label{3.6}
\alpha _1=\beta _1=0\ ,\qquad a_2=b_2=0\ ,
\end{equation}
\begin{equation}\label{3.7}
\alpha _3=\frac{1}{2}Q\,p_0^{-2}(x^2-y^2)\ ,\qquad \beta _3=
Q\,p_0^{-2}x\,y\ ,
\end{equation}
\begin{equation}\label{3.8}
a_3=-2\,D\,x\ ,\qquad b_3=-2\,D\,y\ ,
\end{equation}
\begin{eqnarray}\lefteqn{
c=1-\frac{2m}{r}-2\,\frac{D}{r^2}p_0^{-2}(1-\frac{1}{4}(x^2+y^2))}
\nonumber\\
& & -\frac{Q}{r^3}p_0^{-2}\left\{2-2(x^2+y^2)+\frac{1}{8}(x^2+y^2)^2
\right\}+\dots\ .
\end{eqnarray}
The constant $m$ is interpreted as the mass of the source while 
the constants $D$ and $Q$ are taken to be the dipole and quadrupole 
moments of the source respectively. In $M^+$ the local coordinates 
are $x^\mu _+=(x_+, y_+, r_+, u)$ and 
the multipole moments are $m_+, D_+, Q_+$ etc.. In $M^-$ the local 
coordinates are $x^\mu _-=(x, y, r, u)$ and the multipole moments 
are $m, D, Q$ etc.. The metric tensors  induced on $u=0$ (the boundary 
between $M^+$ and $M^-$) by its embedding in $M^+$ and $M^-$ 
agree provided the following matching conditions are satisfied:
\begin{eqnarray}
x_+&=&x+2\,\frac{[Q]\,x\,p_0^{-1}}{r^3}+\dots\ ,\\
y_+&=&y+2\,\frac{[Q]\,y\,p_0^{-1}}{r^3}+\dots\ ,\\
r_+&=&r+\frac{[Q]p_0^{-2}}{r^2}(x^2+y^2-2)+\dots\ .
\end{eqnarray}
Applying the Barrab\`es--Israel technique yields the results given in 
\cite{BBH} which in the coordinates $(x, y, r)$ read: the 
components of the stress--energy tensor (2.44) 
on $u=0$ are given 
by 
\begin{eqnarray}
16\,\pi\,S^{11}&=&16\,\pi\,S^{22}=-\frac{12\,[Q]}{r^6}(x^2+y^2-2)+\dots\ ,\\
16\,\pi\,S^{12}&=&\frac{24\,[Q]}{r^9}\,p_0^{-2}xy\,(x^2+y^2-2)+\dots\ ,\\
16\,\pi\,S^{13}&=&-\,\frac{6\,[D]\,x}{r^4}+\,\frac{6\,[Q]}{r^5}
\,x\,p_0^{-1}
(x^2+y^2-5)+\dots\ ,\\
16\,\pi\,S^{23}&=&-\,\frac{6\,[D]\,y}{r^4}+\,\frac{6\,[Q]}{r^5}
\,y\,p_0^{-1}
(x^2+y^2-5)+\dots\ ,
\end{eqnarray}
and
\begin{eqnarray}\lefteqn{
16\,\pi\,S^{33}=-\,\frac{4\,[m]}{r^2}+\frac{3\,[D]}{r^3}
\,p_0^{-1}(x^2+y^2-4)}\nonumber\\
& &-\frac{3\,[Q]}{2\,r^4}
\,p_0^{-2}\{24+(x^2+y^2)(x^2+y^2-20)\}+\dots\ .
\end{eqnarray}
The coefficients of the delta function in the Weyl tensor 
are 
\begin{equation}
\hat\Psi _2=-\frac{2\,[Q]}{r^4}\,p_0^{-2}(x^2+y^2-2)+
\dots\ ,
\end{equation}
\begin{equation} 
\hat\Psi _3=\frac{3\,[D]}{\sqrt{2}\,
r^3}\,p_0^{-1}(x-iy)-\frac{3\,[Q]}{\sqrt{2}\,r^4}p_0^{-2}
(x-iy)(x^2+y^2-5)+\dots\ ,
\end{equation}
and 
\begin{equation}
\hat\Psi _4=\,\frac{9\,[Q]}{4\,r^4}
\,p_0^{-2}(x-iy)^2+\dots\ .
\end{equation}
In (3.18)--(3.20) we have an unconventional peeling 
behavior. It is clear now by comparing the conditions (3.6)--
(3.9) and the general formulas (2.56)--(2.59) that {\it this case 
is unconventional because the space--times $M^+$ and 
$M^-$ tend to flatness faster, as future null infinity 
is approached, than the more general space-times $M^+$ and $M^-$ 
considered in section 2}. We note that this signal is free of 
directional singularities. To see unambiguously why $\hat\Psi _4$ 
describes the impulsive gravitational wave part of the signal 
and $\hat\Psi _2$ and $\hat\Psi _3$ describe the light--like 
shell of matter (neutrino burst, for example) the reader 
must consult \cite{BBH}.

\setcounter{equation}{0}
\section{Discussion}\indent
We noted that (2.56)--(2.59) and (3.3) and (3.4) exhibit 
the conventional peeling behavior. Normally one associates 
the radiative part of the field ($\Psi _4$ throughout this 
paper) with a dominant $\frac{1}{r}$--behavior. We see in 
(3.20) that this is not the case in the multipole example. 
On the other hand the radiative part (3.20) of that signal 
is due primarily to the jump in the quadrupole moment of 
the source across the light--like signal and 
this is something that would be expected. The general formulas 
(2.56)--(2.59) allow plenty of scope to construct further 
examples with unconventional peeling behavior. For example we 
could take $M^+$ and $M^-$ to be both Schwarzschild space--
times with masses $m_+$ and $m$. These space--times can 
be attached on $u=0$ asymptotically (for large $r$) with 
the matching (2.27)--(2.30) with $f=x, g=y, h=1$ and 
with (2.34) and (2.35) holding but with $\alpha _1=\beta _1
=0$. Now the stress--energy tensor on $u=0$ is given by 
(2.47)--(2.55) with $a_2=b_2=\alpha _1=\beta _1=0$. Thus 
for example 
\begin{equation}\label{4.1}
S^{33}=-\frac{1}{4\pi\,r^2}\left ([m]-\frac{1}{2}\,h_0\right )+\dots\ ,
\end{equation}
in this case. With $\hat\Psi _A$ calculated from (2.56)--
(2.59) we see that in particular 
\begin{equation}
\hat\Psi _4=\frac{2}{r^2}\frac{\partial}{\partial z}
\left (p_0^2\frac{\partial h_0}{\partial z}\right )+O
\left (\frac{1}{r^3}\right )\ .
\end{equation}
In general this wave will exhibit directional singularities. 
For example if $f_1$ and $g_1$ are constants then
\begin{equation}
\hat\Psi _4=-\frac{1}{8\,r^2}(f_1-ig_1)\,\bar{z}+O\left (\frac{1}{r^3}\right )\ ,
\end{equation}
with $z=x+iy$.
\vskip 2truepc \noindent
{\bf Acknowledgement}\indent
This paper was motivated by a stimulating correspondence 
with Professor J. N. Goldberg. One of us (G. F. B) wishes 
to thank the Department of Education and Science for a 
postdoctoral fellowship.

\end{document}